\documentclass[11pt]{article}
\newcommand{\be}{\begin{equation}}
\newcommand{\ee}{\end{equation}}
\newcommand{\bea}{\begin{eqnarray}}
\newcommand{\eea}{\end{eqnarray}}

\def\({\left(} \def\){\right)}
\usepackage{amssymb}
\usepackage{epsfig}
\usepackage{graphicx}
\begin{document}
\title {A Simplified Mathematical Model for the Formation of Null Singularities Inside Black Holes II} \maketitle
\author{ Dan Gorbonos\footnotemark[1] \ , \
         Gershon Wolansky\footnotemark[2]\\
\footnotemark[1]
 Racah Institute of Physics \\
 The Hebrew University  \\
 Jerusalem 91904  \\
 Israel \\
{\tt gdan@phys.huji.ac.il}

\bigskip
\footnotemark[2] Department of Mathematics \\
  Technion-Israel Institute of Technology \\
  Haifa 32000 \\
  Israel \\
  gershonw@math.technion.ac.il
 \begin{abstract}
We study a simple system of two hyperbolic semi-linear equations,
inspired by the Einstein equations. The system, which was introduced
in~\cite{first}, is a model for singularity formation inside black
holes. We show for a particular case of the equations that the
system demonstrates a finite time blowup. The singularity that is
formed is a null singularity. Then we show that in this particular
case the singularity has features that are analogous to known
features of models of black-hole interiors --- which describe the
inner-horizon instability. Our simple system may provide insight
into the formation of null singularities inside spinning or charged
black holes.
\end{abstract}
\newpage
\tableofcontents

\section{Introduction}

This research examines the singularity formation in a system of
equations motivated by Einstein's equations. This study may lead
to a better understanding of the internal structure of black holes
and more specifically, the singularity created within them. The
system of equations studied in this work is motivated by the
former studies of the Reissner-Nordstr\"{o}m (RN) based
models~\cite{first}. It is a continuation of the work
in~\cite{first} where the system of equations and their physical
motivation were presented. Here we present a new approach to the
study of the model and prove the conjecture presented
in~\cite{first} for a simplified form of the equations.

The mass inflation model presented in~\cite{poisson} demonstrates
 the genericity of null singularities inside realistic
black holes.  In~\cite{first} it was conjectured that a simpler
dynamical system -- an ``active ingredient'' of the Einstein
equations is capable of producing black hole-like configurations of
generic null weak singularities subjected to  certain characteristic
features. It was also shown that the evolution equations in
Reissner-Nordstr\"{o}m-de Sitter (RNDS) spacetime with two
intersecting null fluids can be written in the form:
\begin{eqnarray}
R_{,uv}=e^{s}F(R),  \label{evolution R}\\
s_{,uv}=e^{s}F'(R)  . \label{evolution s}
\end{eqnarray}
 $F(R)$ is defined as:
\begin{equation}
F(R)=-h'(R)\label{definition of F(R)}
\end{equation}
where
\begin{eqnarray}
R\equiv r^{2}, & e^{s}\equiv rf \label{vara},
\end{eqnarray}
for the following metric in double-null coordinates $(u,v)$:
\begin{equation}
ds^{2}=-2f(u,v)dudv+r^{2}(u,v)d\Omega^{2}, \label{metric1}
\end{equation}
and
\[h(R)=2R^{\frac{1}{2}}-4m+\frac{2Q^{2}}{R^{\frac{1}{2}}}-\frac{2\Lambda}{3}R^{\frac{3}{2}},\]
where $m$,$Q$ and $\Lambda$ are the mass, charge and the
cosmological constant in the RNDS solution. The conjecture was
formulated in~\cite{first} for a general function $h(R)$ which has
three roots $h(R_i)=0$ (Fig. $\ref{fig1}$). The roots of $h(R)$
correspond to the locations of the ``horizons'' in a flux-free
solution (defined below). $R=R_{1},R_{2},R_{3}$ correspond in the
RNDS model to the cosmological, event and inner horizons
respectively. The origin of axes is placed at $R=R_{a}$.

In addition we define
\[k_{i} \equiv \left| h'(R_{i})\right| \ \ (i=1,2,3), \] as a
generalization of the surface gravity. In this work we study the
singularity formation for a simple type of $h(R)$ - a saw-tooth
function. We will show that the formed singularity and its
properties depend only on the surface gravities at the horizons and
their relative values.

\begin{figure}[htb]
\vspace{0.5cm} \epsfxsize=0.9\textwidth
\centerline{\epsffile{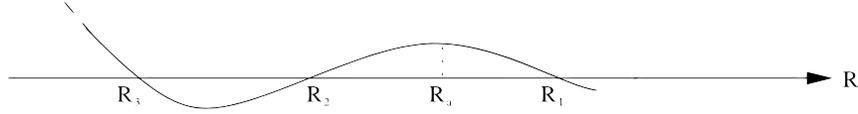}} \vspace*{0.3cm} \caption{\small
\bf \sl The function $h(R)$ - this function can be any function
which has three roots: $R_{1}$, $R_{2}$ and $R_{3}$. $R_{a}$ is the
point where we will place the origin of the axes in the next
sections.} \label{fig1}
\end{figure}
Our semi-linear system is accompanied by two conserved fluxes:
\begin{eqnarray}
e^{s}(e^{-s}R_{,u})_{,u}&=&\Psi(u)\label{constraint u},  \\
e^{s}(e^{-s}R_{,v})_{,v}&=&\Phi(v)\label{constraint v},
\end{eqnarray}
where $\Psi(u)$ and $\Phi(v)$ are arbitrary functions which stand
for the perturbations of the flux-free solution $\Psi(u)=\Phi(v)=0$.
These functions are proportional to the energy density in the
outflux and influx respectively. Eqs. (\ref{constraint
u},\ref{constraint v}) can be derived from our semi-linear system
(see~\cite{first}). Too strong null fluids may ruin the singularity
structure. Thus, we consider ``small'' $\Psi(u)$ and $\Phi(v)$.
Later we will define exactly what we mean by ``small''.

\newtheorem{prop}{Proposition}[section]
\newtheorem{assum}{Assumption}[section]
In addition there exists a gauge freedom ($u \rightarrow
\tilde{u}(u), v \rightarrow \tilde{v}(v)$) for which $s(u,v)$ is
changed according to the following rule:
\begin{equation}
\tilde{s}=s-\ln\left(\frac{d\tilde{v}}{dv}\right)-\ln\left(\frac{d\tilde{u}}{du}\right).
\label{coo}
\end{equation}
There are two quantities which play an important role in the
conjecture and are related to the mass function and the null fluid
fluxes:
\begin{itemize}
  \item $e^{-s}\,R_{,u}\,R_{,v}$ - A gauge invariant quantity.
  \item $e^{-s}\,R_{,u}$ (res. $e^{-s}\,R_{,v}$) - Invariant under
  gauge transformations in $u$ (res. $v$).
\end{itemize}
The paper is organized as follows. In section 2 we study the
singularity formation in the flux-free case using the method of
characteristics. This method will allow us to study the singularity
formation in the non-linear case (with two null fluids -
perturbations of the flux-free case) as well. For this purpose we
introduce in section 3 the general form of characteristic equations.
In the discussion (section 4) we review the conjecture about the
singularity formation (which was presented in~\cite{first}) and
indicate the particular case which will be studied in the rest of
the paper - $h(R)$ as a saw-tooth function. In sections 5 and 6 we
prove the conjecture for a saw-tooth function under few reasonable
assumptions.
 \section{Singularity in the flux-free case and the method of characteristics}
Our semi-linear system ($\ref{evolution R}$,$\ref{evolution s}$)
for the functions $R(u,v)$ and $s(u,v)$ is a system of second
order hyperbolic PDE. To obtain a unique solution inside the
positive quadrate ($u\geq 0,~ v\geq 0$) we should have four known
functions on the boundary of the region: $R(u,0)$, $R(0,v)$,
$s(0,v)$ and $s(u,0)$. This type of initial value problem is
usually called a \textit{Goursat Problem}(see for
example~\cite{Rubinstein}). Notice that the initial data is given
on the characteristic curves of the equations. The Goursat problem
for linear hyperbolic PDE is a well-posed problem and has a global
unique  solution. Our system of equations is semi-linear, so we
have only local existence and uniqueness in general. This local
existence and uniqueness can be proved using successive
approximations as in the linear PDE case.

In addition to the initial data we have a gauge freedom in the
coordinates as our set of equations is invariant under gauge
transformations. We can choose any function for $\frac{\partial
u}{\partial \tilde{u}}$ and $\frac{\partial v}{\partial \tilde{v}}$
(recall the gauge transformation ($\ref{coo}$)) . This freedom
enables us to set a gauge in which $s(0,v)=0$ and $s(u,0)=0$. We
will refer to this gauge later as the ``standard gauge''. In this
gauge we still have to set only two functions: $R(u,0)$ and
$R(0,v)$. Relating to these two functions, we divide the initial
value problem into two cases:
\begin{enumerate}
\item The linear case - $R(u,0)$ and $R(0,v)$ are linear
functions. This is the case when we do not have perturbations
(flux-free).
\item The non-linear case - The two functions $R(u,0)$ and $R(0,v)$
are non-linear.
\end{enumerate}
In the standard gauge, for the flux-free case, eqs. (\ref{constraint
u},\ref{constraint v}) on the boundary $v=0$ and $u=0$ are
\begin{eqnarray}
R_{,uu}=0 & (v=0), \label{linRu}\\
R_{,vv}=0 & (u=0). \label{linRv}
\end{eqnarray}
Therefore, on the $u$ axis $R$ is linear with respect to $u$ and on
the $v$ axis $R$ is linear with respect to $v$. It is consistent
with eqs. ($\ref{linRu}$,$\ref{linRv}$) to assume that $R(u,v)$ is
monotonous on the axes:
\begin{eqnarray}
\left\{\!\!\!\begin{array}{ll} R(u,0)=R_{a}-bu \label{R on u} \\
R(0,v)=R_{a}+av,\label{R on v}
\end{array} \qquad \mbox{$(a,b>0)$}
\right.
\end{eqnarray}
where $R_{a}$ as noted in Fig. $\ref{fig1}$ is placed at the origin
of axes. Equating the expression $e^{-s}R_{,v}R_{,u}$ (which is
gauge invariant) in the static gauge to the same expression in the
standard gauge at the point $R=R_{a}$, we obtain:
\begin{equation}
-h(R_{a})=-e^{-s}R_{,x}^{2}=e^{-s}R_{,v}R_{,u}=-ab. \label{hR0}
\end{equation}

Eqs. (\ref{constraint u},\ref{constraint v}) in the flux-free case
takes the following form:
\begin{eqnarray}
e^{s}(e^{-s}R_{,u})_{,u}&=&0,\label{constraint u a vacuum}  \\
e^{s}(e^{-s}R_{,v})_{,v}&=&0\label{constraint v a vacuum}.
\end{eqnarray}
Our goal is to derive a first order PDE whose characteristic curves
will determine the behavior of the solution of the original
semi-linea system. We start by integrating eqs.($\ref{constraint u a
vacuum}$,$\ref{constraint v a vacuum}$):
\begin{eqnarray}
\left\{\!\!\!\begin{array}{ll} R_{,u}=e^{s}\phi(v) \label{Ru}\\
R_{,v}=e^{s}\psi(u) \label{Rv}
\end{array}
\right.
\end{eqnarray}
 $\phi(v), \psi(u)$ are defined below. After
differentiation, we have:
\begin{eqnarray}
\left\{\!\!\!\begin{array}{ll}
R_{,uv}=e^{s}\phi'(v)+e^{s}s_{,v}\phi(v) \label{1 ruv} \\
R_{,vu}=e^{s}\psi'(u)+e^{s}s_{,u}\psi(u). \label{2 ruv}
\end{array}
\right.
\end{eqnarray}
Since $R_{,uv}$=$R_{,vu}$ ($R(u,v) \in C^{2}$), we obtain a first
order PDE:
\begin{equation}
\phi(v)s_{,v}-\psi(u)s_{,u}+\phi'(v)-\psi'(u)=0. \label{the first
order}
\end{equation}
The characteristic curves for eq. ($\ref{the first order}$) are:
\begin{eqnarray}
\left\{\!\!\!\begin{array}{ll} \dot{v}=-\phi(v) \\
\dot{u}=\psi(u) \\ \label{charac0} \dot{s}=\phi'(v)-\psi'(u).
\end{array}
\right.
\end{eqnarray}
We can find explicitly the functions $\phi(v)$ and $\psi(u)$. In
the standard gauge $s(u,0)=0$ so  (\ref{1 ruv}, \ref{evolution R})
yields
\begin{equation}
R_{,uv}=F(R)=\psi'(u), \label{zpsi}
\end{equation}
\begin{equation}
\psi(u)=\int_{0}^{u}\!F(R(u',0))du'+\psi(0).
\end{equation}
Eqs. ($\ref{R on v}$) and
 ($\ref{Rv}$) in the standard gauge yield:
\begin{eqnarray}
\psi(u)=\int_{R_{a}}^{R}\!F(R)\frac{du'}{dR'}dR'+R_{,v}(0,0)=
-\int_{R_{a}}^{R}\!\frac{F(R')}{b}dR'+a= \nonumber
\\
=\frac{h(R(u,0))-h(R_{a})}{b}+a=\frac{h(R_{a}-bu)}{b}, \label{psi}
\end{eqnarray}
(Here we use the definition of $F(R)$ - eq. ($\ref{definition of
F(R)}$) and the value of $h(R_{a})$ is from eq. ($\ref{hR0}$)).

In the same way, on $u=0$
\begin{eqnarray}
\phi(v)=\int_{R_{a}}^{R}\!F(R)\frac{dv'}{dR'}dR'+R_{,u}(0,0)=
\int_{R_{a}}^{R}\!\frac{F(R')}{a}dR'-b= \nonumber
\\ =-\frac{h(R(0,v))-h(R_{a})}{a}-b=-\frac{h(R_{a}+av)}{a}.
\label{phi}
\end{eqnarray} Now, we can rewrite the characteristic as
\begin{eqnarray}
\left\{\!\!\!\begin{array}{ll} \dot{v}=a^{-1}h(R_{a}+av) \\
\dot{u}=b^{-1}h(R_{a}-bu) \\
\dot{s}=h'(R_{a}-bu)-h'(R_{a}+av). \label{charac}
\end{array}
\right.
\end{eqnarray}

Note the behavior of the function $R(u,v)$ in the region
($u\geq0$,$v\geq0$): From eqs. ($\ref{Ru}$) and ($\ref{phi}$) we
obtain an expression for $R_{,u}$:
\begin{equation}
R_{,u}=-a^{-1}e^{s}h(R_{a}+av).
\end{equation}
We find out (from the form of the function $h(R)$) that $R(u,v)$ is
monotonously decreasing in regions I and II (Fig. $\ref{fig4}$) on
the rays $v=const$ in consistency with the known expression for
$v=0$ (eq. ($\ref{R on u}$)). In an analogous manner, from eqs.
($\ref{Rv}$) and ($\ref{psi}$), we obtain an expression for
$R_{,v}$:
\begin{equation}
R_{,v}=b^{-1}e^{s}h(R_{a}-bu).
\end{equation}
Thus, $R(u,v)$ is monotonously increasing in region I on the rays
$u=const$ in consistency with the known expression for $u=0$ (eq.
($\ref{R on v}$)). In region II $R(u,v)$ is monotonously decreasing
on $u=const$. We summarize the behavior of
 $R(u,v)$ in Fig~\ref{fig4} below.

\begin{figure}[htb]
\vspace{0.5cm} \epsfxsize=0.6\textwidth
\centerline{\epsffile{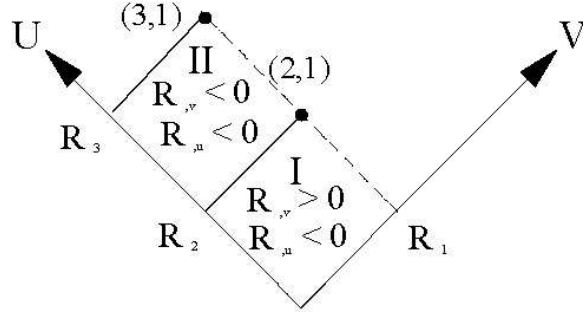}} \vspace*{0.3cm} \caption{\small
\bf \sl The sign of partial derivatives of $R$ in the different
regions of the flux-free solution in the standard gauge. The
indications $R_{1}$, $R_{2}$ and $R_{3}$} label the corresponding
characteristic lines. \label{fig4}
\end{figure}

Next we study the singularity of $s$. As we approach the point (2,1)
in Fig. $\ref{fig4}$, we see that since $h'(R_{2})>0$, $h'(R_{1})<0$
and $h(R)>0$ in region $I$ (R(u,v) is monotonous),
\[\dot{u}, \dot{v} \rightarrow \ 0^{+} \] (the limit is from the positive side of
zero). Thus, $u$ and $v$ attain their asymptotic values. Therefore
from the last equation in (\ref{charac}) we obtain:
\begin{equation}
\dot{s}\rightarrow h'(R_{2})-h'(R_{1})=const>0
\end{equation}
Then, $s\rightarrow \infty$. However, as we approach the point (3,1)
we have two generic cases:
\begin{enumerate}
\item $k_{3}>k_{1}$ - $\dot{s}=const<0$ while $\dot{v}
\rightarrow \ 0^{+}  \Rightarrow s\rightarrow -\infty$. \item
$k_{1}>k_{3}$ - $\dot{s}=const>0$ while $\dot{v} \rightarrow \ 0^{+}
\Rightarrow s\rightarrow \infty$.
\end{enumerate}

{}From (\ref{charac}) we may also consider the flow  along the
characteristics:
\begin{equation}
\frac{du}{dv}=\frac{ah(R_{a}-bu)}{bh(R_{a}+av)}. \label{thuv}
\end{equation}
The direction of this flow is represented in Fig~\ref{fig5}.
\begin{figure}[htb]
\vspace{0.5cm} \epsfxsize=0.6\textwidth
\centerline{\epsffile{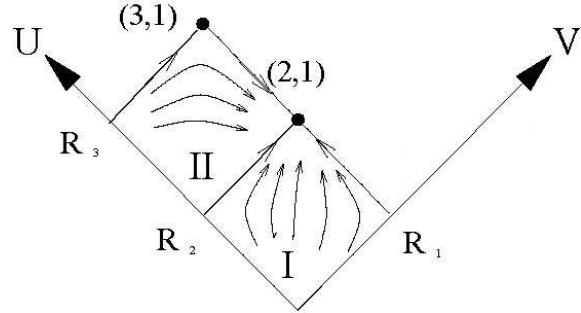}} \vspace*{0.3cm} \caption{\small
\bf \sl The characteristic curves in the flux-free (linear) case in
the standard gauge.} \label{fig5}
\end{figure}

\section{The general characteristic equations}

In this section we derive the characteristic equations in their
most general form, i.e., for any perturbation. Furthermore, we
show that the characteristic curves have physical meaning - they
are the curves of constant $R$. Then we show that the parameter
$t$ of the characteristic curves    also has a physical
interpretation since it is invariant under gauge transformations.

\subsection{Derivation of the characteristic equations}

Similarly to  (\ref{the first order}), we can derive an equation for
$s(u,v)$ for any perturbation to the flux-free case. In this case,
however, the equations are non-linear (and, also, non local) .

Integration of (\ref{constraint u} , \ref{constraint v}) gives:
\begin{eqnarray}
e^{-s}R_{,u}&=&\int^{u}\!e^{-s(u',v)}\Psi(u')du'+\phi(v),
\label{phiv}
\\ e^{-s}R_{,v}&=&\int^{v}\!e^{-s(u,v')}\Phi(v')dv'+\psi(u)
\label{psiu}.
\end{eqnarray}
The  functions $\phi,\psi,\Phi, \Psi$ are determined from $R(0,v)$
and $R(u,0)$ (in the standard gauge) as shown below.
 For convenience, we rewrite
(\ref{phiv},\ref{psiu}) as
\begin{eqnarray}
R_{,u}&=&e^{s}f_{1}(u,v), \label{general Ru} \\
R_{,v}&=&e^{s}f_{2}(u,v) \label{general Rv}.
\end{eqnarray}
Differentiate   (\ref{general Ru}) with respect to $v$,
(\ref{general Rv}) with respect to $u$, and equate
$R_{,uv}=R_{,vu}$ to obtain
\begin{equation}
s_{,v}f_{1}-s_{,u}f_{2}+f_{1,v}-f_{2,u}=0. \label{form}
\end{equation}
The equations of the characteristic curves for this equation are:
\begin{eqnarray}
\left\{\!\!\!\begin{array}{ll} \dot{v}=-f_{1}(u,v) \\
\dot{u}=f_{2}(u,v) \\ \label{gc} \dot{s}=f_{1,v}-f_{2,u}.
\end{array}
\right.
\end{eqnarray}
{}From  (\ref{general Ru}, \ref{general Rv}) and (\ref{gc}), we
obtain:
\begin{equation}
\frac{du}{dv}=-\frac{f_{2}}{f_{1}}=-\frac{R_{,v}}{R_{,u}} \ .
\label{bit}
\end{equation}
It follows
\begin{prop}
A curve in the domain $u \geq 0$ $v \geq 0$ is a characteristic
curve of ($\ref{form}$) if and only if $R=const$ on the curve.
\end{prop}

 As in the
linear case, we can obtain from (\ref{phiv}-\ref{general Rv}):
\begin{equation}
R_{,uv}=e^{s}s_{,v}(\int^{u}_{0}\!e^{-s(u',v)}\Psi(u')du'+\phi(v))-e^{s}(\int^{u}_{0}\!s_{,v}e^{-s}\Psi(u')du'-\phi'(v))
\label{general zphi}
\end{equation}
\begin{equation}
R_{,vu}=e^{s}s_{,u}(\int^{v}_{0}\!e^{-s(u,v')}\Phi(v')dv'+\psi(u))-e^{s}(\int^{v}_{0}\!s_{,u}e^{-s}\Psi(v')dv'-\psi'(u)).
\label{general zpsi}
\end{equation}
{}From  (\ref{general zphi}) on $u=0$ in the standard gauge
($s(0,v)=0$) we obtain  $R_{,uv}=\phi'(v)$.  From  (\ref{general
zpsi}) on $v=0$  in the standard gauge ($s(u,0)=0$) we obtain
$R_{,vu}=\psi'(u)$. Equating these expressions to eq.
($\ref{evolution R}$), we arrive at the same results for the
derivatives of $\psi(u)$ and $\phi(v)$ as in the linear case:
\begin{equation}
-h'(R(u,0))=F(R(u,0))=\psi'(u) \label{zpsi gen}
\end{equation}
\begin{equation}
-h'(R(0,v))=F(R(0,v))=\phi'(v). \label{zphi gen}
\end{equation}

Note also that in the standard gauge the conserved fluxes
(\ref{constraint u}, \ref{constraint v}) takes, on the axis $u=0$,
$v=0$, the form:
 \bea \Psi(u)=R_{,uu}(u,0)\\
\Phi(v)=R_{,vv}(0,v). \eea Combining the above results we can
rewrite the characteristic equations (\ref{gc}) explicitly (in
terms of the data $R(u,0)$ and $R(0,v)$), as
\begin{eqnarray}
\left\{\!\!\!\begin{array}{ll} \label{general charac new}
\dot{v}=-\int^{u}\!e^{-s}\,R_{,uu}(u',0)du'+\int^{v}\!h'\(R(0,v')\)\,dv'\\
\dot{u}=\int^{v}\!e^{-s}\,R_{,vv}(0,v')dv'-\int^{u}\!h'\(R(u',0)\)\,du' \\
\dot{s}=h'(R(u,0))-h'(R(0,v))+\int^{u}\!\(e^{-s}\)_{,v}\,R_{,uu}(u',0)du'-\int^{v}\!\(e^{-s}\)_{,u}\,R_{,vv}(0,v')dv'.
\end{array}
\right.
\end{eqnarray}

\subsection{The physical interpretation of t}

The parameter $t$ of the characteristic curves is related to the
proper time in the same way as the coordinates $r^{*}$ or $t$ are
related to it\footnote{$t$ here is the parameter that we use in the
usual black-hole metric, $r^{*}$ is the \textit{tortoise
coordinate}, defined by $r^{*}=\int\! \frac{dr}{1-\frac{2m}{r}}$ for
example in the Schwarzschild metric.}. The proper time $\tau$ is
defined by:
\begin{equation}
d\tau^{2}\equiv\frac{2e^{s}}{R^{\frac{1}{2}}}dudv.
\end{equation}
{}From eqs. ($\ref{gc}$) and ($\ref{general Ru}$) we obtain:
\begin{equation}
\dot{u}=R_{,v}e^{-s}. \label{dudt}
\end{equation}
{}From (\ref{general Rv}) and (\ref{gc}) we obtain:
\begin{equation}
\dot{v}=-e^{-s}R_{,u}. \label{dvdt}
\end{equation}
Combining the last three results, we find that the proper time
$\tau$ and the parameter $t$ are connected through the scalar
$e^{-s}R_{,u}R_{,v}$:
\begin{equation}
\left(\dot{\tau}\right)^{2}=\left|
\frac{2e^{-s}R_{,v}R_{,u}}{R^{\frac{1}{2}}} \right|.
\end{equation}
Therefore the parameter $t$ is a geometric quantity and not
coordinate dependent. In the linear case we have:
\begin{equation}
\left(\dot{\tau}\right)^{2}=\left|\frac{2h(R)}{R^{\frac{1}{2}}}\right|.
\label{rel}
\end{equation}
It appears to be like the connection between $\tau$ and $t$ on
$r=const$, where $t$ is the usual Schwarzschild coordinate (recall
that $\frac{h(R)}{2R^{\frac{1}{2}}}=2f$ in the static solution), but
here the characteristic curves are more general. For example, the
relation ($\ref{rel}$) is valid for curves of the type $u=const$ and
$v=const$ even though it is not true for the Schwarzschild $t$
coordinate.  Such curves are the characteristics along the
``horizons''. Furthermore, on each characteristic curve $R=const$
and therefore $\left(\dot{\tau}\right)^{2}=const$. Thus, the proper
time is proportional to the parameter $t$, except for the horizons
(roots of $h(R)$) along which $\dot{\tau}=0$). To conclude, in the
linear case the parameter $t$ is regarded as an extension of the
Schwarzschild coordinate $t$ to include the horizons as well.

\section{Discussion}
\label{the discussion} Let us summarize the features of the
singularity obtained in the linear case
($R_{,uu}(u,0)=R_{,vv}(0,v)\equiv0$). We have a singularity that
develops from regular initial conditions. This singularity forms at
finite $u$ and $v$ - a finite time blowup. Its structure suits the
structure of the RNDS spacetime (with an additional ``horizon''
because of the additional zero in $h(R)$). The singularity consists
of a ``point singularity'' where $s\rightarrow\infty$ and a ray of
singularity emerging from it where $s\rightarrow\pm\infty$,
depending on inequalities between the derivatives of $h(R)$ at the
horizons. The structure is illustrated in Fig. $\ref{fig14}$. These
inequalities and the resulting singularities of  $s$ correspond to
the blue-shift$\backslash$red-shift at the inner horizon.
\begin{figure}[htb]
\vspace{0.5cm} \epsfxsize=0.6\textwidth
\centerline{\epsffile{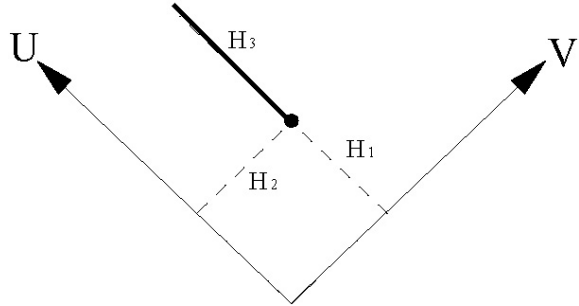}} \vspace*{0.3cm} \caption{\small
\bf \sl The general singularity structure. The three horizons are
displayed, denoted by $H_{i}$ ($i=1,2,3$). The singular inner
horizon is displayed by a thick line emerging from the point
singularity in the middle.} \label{fig14}
\end{figure} In addition,
we found out that each horizon corresponds to a certain $R$ value
(which is constant on the horizon); In the linear case the horizons
are characteristic curves of the first order PDE ($\ref{the first
order}$) according to Proposition 3.1. In the non-linear case  the
curves of constant $R$  deviate from the horizons due to the
perturbations. Therefore we will use in the non-linear case the
notation $H_{i}$ ($i=1,2,3$) for the horizons. $H_{1},H_{2},H_{3}$
correspond in the RNDS model to the cosmological, event and inner
horizons respectively.

We reformulate below the main ingredient of the
conjecture\footnote{The full form of the conjecture can be found
in~\cite{first}} for the non-linear case (nonzero flux in both
directions) :
\begin{center}{\bf Conjecture}\end{center}
\emph{For small enough non-linear perturbations the  structure of
the linear case is preserved. That is, a singularity point is formed
from which the inner horizon emerges ($H_3$ in Fig.~\ref{fig14}). In
addition, we expect the gauge invariant quantities to diverge at the
inner horizon. Since they are related to the mass inflation
phenomenon mentioned in the introduction, we expect the conditions
for their divergence to be the same as in the mass inflation model
of Brady and Poisson~\cite{Brady2}.}
\par
At the table below we summarize the divergence conditions of
different quantities at the inner horizon based on~\cite{first} for
the linear and non-linear cases.

\begin{table}[
htb] \centering \vspace*{0.3cm}
\begin{tabular}{|c||c|c|c|} \hline
 {\emph{The quantity}} & \multicolumn{2}{c | }{\emph{Type of divergence}} & \emph{Condition for}       \\
  &\multicolumn{2}{c | }{} & \emph{the divergence}
 \\ \hline \hline
  & linear & non-linear & \\
  \cline{2-4}
$s$ & $+\infty$ & $+\infty$ & $k_{1}>k_{3}$ \\
    & $-\infty$ & $-\infty$ & $k_{3}>k_{1}$ \\ \hline
$e^{-s}R_{,v}$ & --- & $+\infty$ & $k_{3}>2k_{1}$ \\
\hline $e^{-s}R_{,v}R_{,u}$ & --- & $+\infty$ & $k_{3}>2k_{1}$
\\ \hline
\end{tabular}
\caption{\small \bf \sl The divergence of various quantities at the
inner horizon $H_{3}$ expected to be found in the linear and
non-linear cases.}
\end{table}

\subsection*{ $h(R)$ as a saw-tooth function}

{}From now on we write the characteristic equations for a particular
case of the function $h(R)$. We choose $h(R)$ to be a saw-tooth
function as illustrated in Fig. $\ref{fig6}$. The origin of the
$(u,v)$ axes is positioned  on the curve $R=R_{a}$.  The region
where $R_{a}<R(u,v)<R_{1}$ will be referred as region 1. The region
where $R_{b}<R(u,v)<R_{a}$ is region 2. These, and regions 3 and 4
are displayed in Fig.~\ref{regions}.

\begin{figure}[htb]
\vspace{0.5cm} \epsfxsize=\textwidth
\centerline{\epsffile{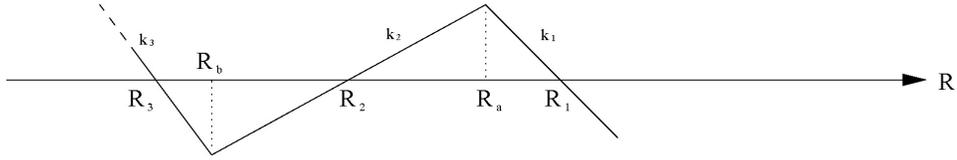}} \vspace*{0.3cm} \caption{\small
\bf \sl The function h(R) as a ``saw-tooth'' function.} \label{fig6}
\end{figure}

\begin{figure}[htb]
\vspace{0.5cm} \epsfxsize=0.6\textwidth
\centerline{\epsffile{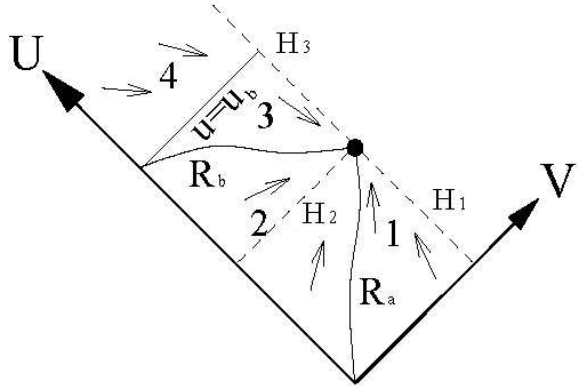}} \vspace*{0.3cm} \caption{\small
\bf \sl The regions 1,2,3,4 in the solution -  the forms of these
regions are determined by the form of the function $h(R)$. The
general directions of the characteristic curves in each region are
indicated by arrows.} \label{regions}
\end{figure}

For the saw-tooth function the evolution equation for $s$
(\ref{evolution s}) turns to be in each region $s_{,uv}=0$ whose
solution is
\begin{equation}\label{u+v}s(u,v)=s_{1}(u)+s_{2}(v),\end{equation} where
$s_{1}(u)$ is an arbitrary function of $u$ and $s_{2}(v)$ - a
function of $v$.

\section{Reduction to a system of ODE}

In this section we reduce our semi-linear system to a system of four
first order ODE along a characteristic curve.
 First we begin the discussion by investigating the characteristic curve
$R(u,v)=R_{a}$. We choose the parameter $t$ along this
characteristic curve to be the independent variable for the
equations. Our semi-linear system (\ref{evolution R},\ref{evolution
s}) takes the following form near $R(u,v)=R_{a}$:
\begin{eqnarray}
\left\{\!\!\!\begin{array}{ll}
R_{,uv}=e^{s}[(k_{1}+k_{2})\theta(R-R_{a})-k_{2}] \ \label{evol1}\\
s_{,uv}=e^{s}(k_{1}+k_{2})\delta(R-R_{a}), \label{evol2}
\end{array}
\right.
\end{eqnarray}
where $\theta(s)$ is the Heaviside step function ($\theta(s)=1$ for
$s\geq 0$ and $\theta(s)=0$ for $s<0$). These equations are valid in
the neighborhood of $R=R_{a}$. In the standard gauge $s=s_{1}(u)$ is
the solution in region 1 and $s=s_{2}(v)$ is the solution in region
2 - see (\ref{u+v}). \footnote{We assume here that there is no
``hiding'' of the singularity --- namely, the curve $R_{b}$ is
monotonously decreasing (Fig. $\ref{regions}$). This assumption is
included in the statement that $s$ is only a function of $v$. In the
next section we will formulate sufficient conditions for it.} On the
curve $R=R_{a}$:
\begin{equation}
s_{1}(u)=s_{2}(v).
\end{equation}
Eq. ($\ref{evol1}$) takes the form $R_{,uv}=k_{1}e^{s_{1}(u)}$ in
region 1. We integrate this equation with respect to $u$:
\begin{equation}\label{83}
R_{,v}=\int^{u}\!k_{1}e^{s_{1}(u')}du'+R_{,v}(0,v),
\end{equation}
and then differentiate with respect to $t$. Using
(\ref{dudt},\ref{dvdt}) we obtain the following equation for
$R_{,v}$:
\begin{equation}
\dot{R}_{,v}=k_{1}\,R_{,v}-R_{,vv}(0,v)\,e^{-s}\,R_{,u}.
\end{equation}
We obtain an equation for $R_{,u}$ by integrating eq.
($\ref{evol1}$) in region 2 with respect to $v$, and then
differentiate with respect to $t$ (using again
(\ref{dudt},\ref{dvdt})):
\begin{equation}
\dot{R}_{,u}=k_{2}\,R_{,u}+R_{,uu}(u,0)e^{-s}R_{,v}. \label{e2}
\end{equation}
We obtain the third equation by integrating eq. ($\ref{evol2}$) with
respect to $u$ in region 2 (in the standard gauge $s_{,v}(0,v)=0$):
\begin{equation}
s_{2}'(v)=s_{,v}=(k_{1}+k_{2})\int^{u}_{0}\!
e^{s(u',v)}\delta(R-R_{a})du'=(k_{1}+k_{2})
\frac{e^{s_{2}(v)}}{\left| R_{,u}(u_{0},v) \right|},
\end{equation}
where $u_{0}$ is the solution of $R(u_{0},v)=R_{a}$. Then using
(\ref{dvdt})we have \be \dot{s}=-(k_{1}+k_{2})\,\frac{R_{,u}}{\left|
R_{,u}(R_{a}) \right|}, \ee
 where we denote by
$R_{,u}(R_{a})$ the function $R_{,u}$ on the curve $R(u,v)=R_{a}$.
Note that the equation for $s$ is written in a form which is valid
in all of region 2. When we add (\ref{dudt}) to the above three
equations we get the complete system of ODE on the curve $R=R_{a}$:
\begin{eqnarray}
\left\{\!\!\!\begin{array}{ll}
\dot{R}_{,v}=k_{1}\,R_{,v}-R_{,vv}(0,v)\,e^{-s}\,R_{,u}\\
\dot{R}_{,u}=k_{2}\,R_{,u}+R_{,uu}(u,0)e^{-s}R_{,v} \\
\dot{s}=-(k_{1}+k_{2})\,\frac{R_{,u}}{\left| R_{,u}(R_{a})\right|} \\
\dot{u}=e^{-s}\,R_{,v}.
\end{array}
\right. \label{the4}
\end{eqnarray}
Our goal now is to extend this set of four equations to any
characteristic curve in region 2. It is only the first equation
which has to be replaced in region 2.  For this  we recall the
characteristic equations (\ref{general charac new}) and equate the
equation for $u$ along the characteristic with (\ref{dudt}). After
differentiation with respect to $t$ and using (\ref{dudt},
\ref{dvdt}) we obtain  \be
\dot{R}_{,v}=-k_{2}\,R_{,v}-(k_{1}+k_{2})\,\frac{R_{,u}}{\left|
R_{,u}(R_{a})\right|}-R_{,vv}(0,v)\,e^{-s}\,R_{,u}. \ee

This set of four equations is accompanied by initial conditions on
the $u$ axis for $(R_{,v},R_{,u},s,u)$ derived from our original
semi-linear system. For a characteristic curve in the standard gauge
originated at $(\overline{u},0)$ we obtain:
\begin{eqnarray}
\left\{\!\!\!\begin{array}{ll}
R_{,v}(0)=R_{,v}(0,0)-k_{2}\overline{u}
\\ R_{,u}(0)=R_{,u}(\overline{u},0)
\\ s(0)=0
\\ u(0)=\overline{u}.
\end{array}
\right. \label{initial}
\end{eqnarray}

In regions 3 and 4 we shall only need  the equation for the function
$s$. By the same argument as in the previous discussions, $s(u,v)$
is only a function of $v$ in region 4 (in the standard gauge). We
denote $s$ in region 4 by $s_{4}(v)$. In region 3 we can write the
solution as $s(u,v)=s_{4}(v)+s_{3}(u)$, where $s_{4}(v)$ is the
solution in region 4 (by continuity). Now we write our semi-linear
system (\ref{evolution R}, \ref{evolution s}) with the relevant
terms in the neighborhood of the curve $R=R_{b}$:
\begin{eqnarray}\label{xxx}
\left\{\!\!\!\begin{array}{ll}
R_{,uv}=e^{s}[-(k_{2}+k_{3})\theta(R-R_{b})+k_{3}] \\
s_{,uv}=e^{s}[(k_{1}+k_{2})\delta(R-R_{a})-(k_{2}+k_{3})\delta(R-R_{b})].
\end{array}
\right.
\end{eqnarray}

In regions 3 and 4 we have to integrate the second of (\ref{xxx})
with respect to $u$ over the delta functions  across $R=R_{a}$ and
$R=R_{b}$ and use the first equation in (\ref{xxx}) to obtain:
 \be
s_{4}'(v)= e^{s_{2}(v)}\(\frac{(k_{1}+k_{2})}{\left|
R_{,u}(R_{a})\right|}-\frac{(k_{2}+k_{3})}{\left|
R_{,u}(R_{b})\right|}\), \label{s3non} \ee where  $R_{,u} (R_{a})$
is the function $R_{,u}(v)$ on the curve $R=R_{a}$ and
$R_{,u}(R_{b})$ is the function $R_{,u}(v)$ on the curve $R=R_{b}$.

\section{The non-linear case}

\subsection{Singularity formation in the non-linear case}
The structure  of Fig~\ref{regions}, where $R_a$ is the boundary
between regions 1 and 2, implies $R_{,u}(0,0)<0$. Therefore there
exists $\tilde{t}>0$, so that $R_{,u}<0$ on $R=R_{a}$ for $0\leq
t<\tilde{t}$. On this segment, (\ref{the4}) holds while the third
equation is simplified into $\dot{s}=k_1+k_2$. Then in the standard
gauge we obtain:
 \be
s=(k_{1}+k_{2})\,t. \label{thes}\ee

\begin{prop} \label{thesing}
Assume $m_{1}<R_{,v}(0,v)<M_{1}$ and $-M_{2}<R_{,u}(u,0)<-m_{2}$
where $m_{1},M_{1},m_{2},M_{2}$ are arbitrary positive constants.
Then, there exists singularity $s\rightarrow\infty$  on the curve
$R=R_{a}$ at  finite values of $u$ and $v$.

\end{prop}
\textit{Proof}\\ Using eqs. (\ref{dvdt}, \ref{dudt}) and the first
two equations in (\ref{the4}) we obtain
\begin{eqnarray}\label{94}
\left\{\!\!\!\begin{array}{ll}
R_{,v}(t)=k_{1}e^{k_{1}t}\int^{t}_{0}\! e^{-k_{1}t'}R_{,v}(0,v)dt'+R_{,v}(0,v) \\
R_{,u}(t)=k_{2}e^{k_{2}t}\int^{t}_{0}\! e^{-k_{2}t'}R_{,u}(u,0)dt'+R_{,u}(u,0). \\
\end{array}
\right.
\end{eqnarray}
{}From this and the assumption of the proposition on $R_{,v}(0,v)$
and $R_{,u}(u,0)$:
\begin{eqnarray}
\left\{\!\!\!\begin{array}{ll}
m_{1}e^{k_{1}t}<R_{,v}(t)<M_{1}e^{k_{1}t}
\\ \label{inequa}
-M_{2}e^{k_{2}t}<R_{,u}(t)<-m_{2}e^{k_{2}t}.\\
\end{array}
\right.
\end{eqnarray}
 Now we substitute (\ref{thes}) into
(\ref{dvdt}, \ref{dudt}) to obtain
\begin{eqnarray}
\left\{\!\!\!\begin{array}{ll}
\dot{u}=e^{-(k_{1}+k_{2})t}R_{,v}\\
\dot{v}=-e^{-(k_{1}+k_{2})t}R_{,u},  \\
\end{array}
\right.
\end{eqnarray}
which, together with  (\ref{inequa}), implies
\begin{eqnarray}
\left\{\!\!\!\begin{array}{ll}
u(t)<\frac{M_{1}}{k_{2}}(1-e^{-k_{2}t})\\
v(t)<\frac{M_{2}}{k_{1}}(1-e^{-k_{1}t}). \\
\end{array}
\right.
\end{eqnarray}
In conclusion, when $t\rightarrow\infty$, $s\rightarrow\infty$ but
$\lim_{t\rightarrow\infty}u(t)<\frac{M_{1}}{k_{2}}$ and
$\lim_{t\rightarrow\infty}v(t)<\frac{M_{2}}{k_{1}}$. Thus, the
singularity forms at a finite $u$ and $v$. $\Box$\\

Under the assumption of Proposition~\ref{thesing} we obtain from
(\ref{inequa}) that $R_{,u}(t)<0$ for $t\in[0,\infty)$ on $R=R_{a}$.
The same conclusion is valid for any characteristic curve in region
2 since the equation for $R_{,u}$ takes the same form as on
$R=R_{a}$.

In Fig. \ref{regions} we assume that the curve $R=R_{b}$ is
monotonously decreasing. This is guaranteed under the following
reasonable assumption
\begin{assum}
$R_{,vv}(0,v)<0$ (the weak energy condition) \label{weak}
\end{assum}

 The characteristic curve which plays an important role in
determining whether $R_{b}$ is monotonous or not is the curve
which starts on the $u$ axis at the point
$u_{h}=k_2^{-1}R_{,v}(0,0)$. We denote the constant value of $R$
on this curve by $R_{h}$. In the next result  we show that
$R=R_{h}$ is monotonously decreasing, and that its $u_h$ is the
minimal value of $u$ with this property.

\begin{prop}
Let Assumption~\ref{weak}.
  The
characteristic curve originated at $u_{h}=k_2^{-1}R_{,v}(0,0)$ on
the $u$ axis is monotonously
decreasing ($\dot{u}<0$) provided $u_{b}>u_{h}$ (see Fig. \ref{regions}).\\
\label{propmon}
\end{prop}
\textit{Proof}  Let us write the characteristic equation for
$u(t)$ (\ref{general charac new}) in region 2:
\[
\dot{u}=\int_{0}^{v}\! e^{-s}R_{,vv}(0,v')dv'+R_{,v}(0,0)-k_{2}u.
\]
Even though we still do not know the shape of the curve $R=R_{h}$
, we know that in a small region near the axis $u$, $s$ is a
function only of $v$ (a consequence of the choice in the standard
gauge). This region is shown in Figure~\ref{fig9}.  There exists
$0<t_{f}$ such that for any
 $t\in[0,t_{f}]$, $s=s(v)$. Now we differentiate
this equation with respect to $t$ and use (\ref{dvdt}) to obtain
\begin{equation}
\ddot{u}+k_{2}\dot{u}+e^{-2s(v)}R_{,vv}(0,v)R_{,u}=0 \label{Weq}
\end{equation}
By the choice $u_{h}=k_2^{-1}\,R_{,v}(0,0)$ we have
$\dot{u}(0)=0$. Next we  show that $\dot{u}(t)<0$ for any
$t\in(0,t_{f})$. Substituting $\dot{u}(t)=c(t)e^{-k_{2}t}$ into
(\ref{Weq}) we find
\begin{equation}
\dot{c}(t)e^{-k_{2}t}=-e^{-2s}R_{,vv}(0,v)R_{,u}. \label{thec}
\end{equation}
Since $R_{,vv}(0,v)<0$ (assumption \ref{weak} - the weak energy
condition) and $R_{,u}<0$ in region 2 ($\dot{v}=-e^{-s}R_{,u}>0$
in region 2 from  (\ref{dvdt})), we find that $\dot{c}(t)<0$. In
addition, $c(0)=\dot{u}(0)=0$. Therefore $\dot{u}<0$ for any
$t\in(0,t_{f})$. Note that $\dot{v}>0$ in region 2 where $s=s(v)$,
thus $\frac{du}{dv}<0$ on the characteristic curve $R=R_{h}$.

\begin{figure}[htb]
\vspace{0.5cm} \epsfxsize=0.5\textwidth
\centerline{\epsffile{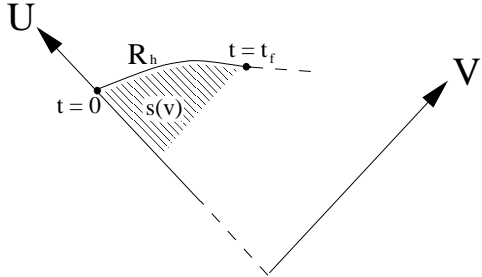}} \vspace*{0.3cm} \caption{\small \bf
\sl The beginning of the curve $R=R_{h}$ (the curve that starts at
the point $u_{h}=k^{-1}_{2}\,R_{,v}(0,0)$). Here we assume that the
curve starts monotonously decreasing. According to Proposition
\ref{propmon} this is the correct possibility.} \label{fig9}
\end{figure}
We now argue that $t_f=\infty$. Indeed, if $t_f<\infty$ then,
necessarily, $\dot{u}(t_f)=0$. This, however, is impossible via
(\ref{thec}), whose right side is negative in region~2.
$\Box$\\
\begin{figure}[htb]
 \vspace{0.5cm}
\epsfxsize=0.5\textwidth
\centerline{\epsffile{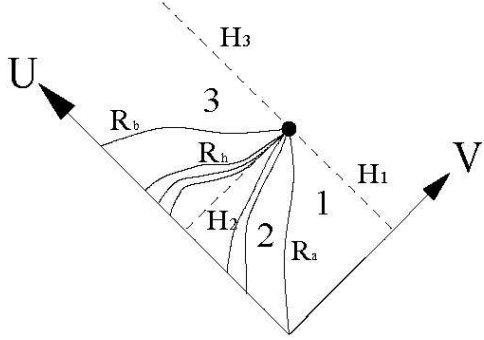}} \vspace*{0.3cm} \caption{\small
\bf \sl The curve $R=R_{h}$ and some other characteristic curves.
Notice that the curve $R=R_{h}$ is the first characteristic curve
which is monotonously decreasing (first in the sense of minimal $u$
values) . The characteristic curves before $R=R_{h}$ are not
monotonous.} \label{fig11}
\end{figure}

The characteristic curve originated at   $(u_{h},0)$ plays an
important role in defining the boundary of region 2. If this point
is part of the boundary of region 2, then the singularity is not
hidden.   Based on the assumption that $-M_{2}<R_{,u}(u,0)<-m_{2}$
 we obtain:
\[R_{a}-M_{2}u<R(u,0)<R_{a}-m_{2}u.\]
Thus we can bound $u_{b}$ from below:
\begin{equation}
u_{b}>\frac{R_{a}-R_{b}}{M_{2}}.
\end{equation}
Therefore, a sufficient condition for $u_{b}\geq
u_{h}(=k_2^{-1}R_{,v}(0,0))$ in the non-linear case is
\begin{equation}\label{9.3}
R_{a}-k_2^{-1}M_{2}R_{,v}(0,0)>R_{b}.
\end{equation}
\newtheorem{lem}{Lemma}[section]
\begin{prop}
All the characteristic curves in region 2 meet at one point where
$s\rightarrow\infty$.
\end{prop}
\textit{Proof}\\ First we argue that any characteristic curve $R
\neq R_{a}$ cannot intersect with $R_{a}$ at $t<\infty$. If such
an intersection takes place, then the gradient of $R$ will blow
up. The latter is impossible for $t<\infty$ according to
(\ref{inequa}). The same argument applies for any two
characteristic curves in region 2.  We conclude that the
characteristic curves in region 2 cannot intersect for any
$t<\infty$.

We showed in Proposition \ref{thesing} that the singularity exists
on the curve $R=R_{a}$. Let us denote the point singularity on the
curve $R=R_{a}$ by $(u_{s},v_{s})$. Since in region 2, $s(u,v)$ is a
function only of $v$, we have the locus of the singularities in
region 2:
$$R^{\infty}(u_0)\equiv \lim_{t\rightarrow\infty} R(t) \ \ ; \ \
 v(0)=s(0)=0, u(0)=u_0, R(0)= R(u_0,0),$$ where $0\leq u_{0}\leq u_{b}$
(see Fig.~\ref{fig12} where the locus and one of the characteristic
curves $R=\tilde{R}$ are displayed). On $R^{\infty}(u)$ we have a
divergence of $s$ via (\ref{thes}).
 Since no two
characteristic curves can intersect at $t<\infty$ in region 2, then
$R^{\infty}(u_1)\geq R^\infty(u_2)$ if $0\leq u_1 < u_2 \leq u_b$.
If, for some $u_1<u_2$, $R^{\infty}(u_1)> R^\infty(u_2)$, then
$$\int_{u_1}^{u_2}\! R^{\infty}_{,u} du
=R^{\infty}(u_2)-R^{\infty}(u_1)<0 \ , $$ which implies that $ \left
|\!R^\infty_{,u}\!\right| < \infty$ for almost any $u$ in the
interval $(u_1,u_2)$. But $ R^\infty_{,u} =
\lim_{t\rightarrow\infty} R_{,u}(t)=\infty$ (from (\ref{inequa}))
where $R_{,u}(t)$ is given by (\ref{94}), and we obtain a
contradiction.  It follows that $R^\infty$ is a constant in any such
interval.   $\Box$
\begin{figure}[htb] \vspace{0.5cm}
\epsfxsize=0.6\textwidth
\centerline{\epsffile{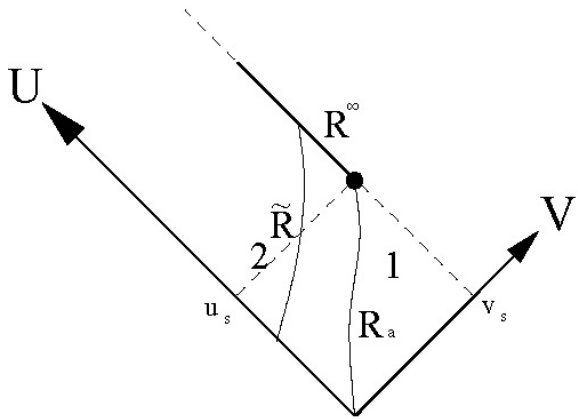}} \vspace*{0.3cm} \caption{\small
\bf \sl The curve $R=\tilde{R}$ and the curve $R=R_{a}$ hit the ray
of singularity ($R^{\infty}$). The situation  in this figure is
impossible as we show in the proof of Proposition 8.3} \label{fig12}
\end{figure}

In order to proceed to regions 3 and 4, we need another result about
region 2 which is formulated in the following Lemma.

\begin{lem}
In region 2 $\frac{R_{,u}(R')}{R_{,u}(R'')}\rightarrow1$ when
$t\rightarrow\infty$, for any $R',R''\in[R_{a},R_{b}]$
\end{lem}
\textit{Proof}\\
{}From equation (\ref{evol1}) we obtain, analogously to (\ref{83}),
$$ R_{,u}=-k_2\int^v \! e^{s_2(v^{'})} dv^{'} + R_{,u}(u,0) \ . $$
This equation  is valid for any characteristic curve in region 2.
So,  look at the limit
\begin{equation}
\lim_{v\rightarrow
v_{s}}\frac{R_{,u}(R')}{R_{,u}(R'')}=\lim_{v\rightarrow
v_{s}}\frac{-k_{2}\int^{v}\!e^{s_{2}(v')}dv'+R_{,u}(u'(v),0)}{-k_{2}\int^{v}\!e^{s_{2}(v')}dv'+R_{,u}(u''(v),0)}.
\end{equation}
$u'(v)$ and $u''(v)$ describe the curves $R(u,v)=R'$ and
$R(u,v)=R''$ respectively. According to (\ref{thes}) the integrals
in the denominator and the numerator diverge. Using L'Hospital's
rule we obtain:
\begin{equation}
\lim_{v\rightarrow
v_{s}}\frac{R_{,u}(R')}{R_{,u}(R'')}=\lim_{v\rightarrow
v_{s}}\frac{-k_{2}+R_{,uu}(u'(v),0)e^{-s_{2}(v)}\frac{du'}{dv}}{-k_{2}+R_{,uu}(u''(v),0)e^{-s_{2}(v)}\frac{du''}{dv}}.
\end{equation}
$\lim_{v\rightarrow v_{s}}e^{-s_{2}(v)}\frac{du}{dv}=0$ on any
characteristic because $\lim_{v\rightarrow
v_{s}}\int^{v}\!e^{s_{2}(v')}dv'$ diverges but $\lim_{v\rightarrow
v_{s}}\int^{v}\!\frac{du}{dv'}dv'=u(v_{s})$ is finite. Therefore
$\lim_{v\rightarrow v_{s}}\frac{R_{,u}(R')}{R_{,u}(R'')}=1$.
$\Box$

Let us summarize the properties of the inner horizon of our
toy-model in the following proposition.
\begin{prop}
Assume (\ref{9.3}). Then, at the  inner horizon $H_{3}$:\\
a.  If $k_{3}>k_{1}$, $s\rightarrow\infty$, while if $k_{1}<k_{3}$, $s\rightarrow-\infty$.\\
b.  $R_{,v}e^{-s}$ diverges if $k_{3}>2k_{1}$ and converges if $k_{3}<2k_{1}$. \\
c. $e^{-s}R_{,v}R_{,u}$ diverges if $k_{3}>2k_{1}$ and converges
if $k_{3}<2k_{1}$.
\end{prop}
\textit{Proof}

 a.
 In region 4 we have an expression for $s_{4}(v)$ as a function of
$s_{2}(v)$ and $R_{,u}$ in region 2 (eq. ($\ref{s3non}$)):
\begin{equation}
s_{4}(v)=\int^{v}_{0}\!\frac{1}{R_{,u}(R_{b})} \left[
k_{2}+k_{3}-\frac{(k_{1}+k_{2})\,R_{,u}(R_{b})}{R_{,u}(R_{a})}
\right] e^{s_{2}(v')}dv'. \label{ss3}
\end{equation}

Let us look at the integral when $v\rightarrow v_{s}$. The integrand
is written in terms of quantities in region 2. Recall that in region
2 $R_{,u}<0$. Then since $s_{2}(v)\rightarrow\infty$ when
$v\rightarrow v_{s}$ and using Lemma 6.1 we see that if
$k_{3}>k_{1}$ then $s_{4}(v)\rightarrow-\infty$ and if
$k_{1}>k_{3}$, $s_{4}(v)\rightarrow\infty$. Note that $s_{4}(v)$ is
regular in all region 4 for $v<v_{s}$.

In region 3, although $s(u,v)=s_{4}(v)+s_{3}(u)$, we have the same
type of singularity as in region 4 along the horizon $H_{3}$ (except
for the point singularity). $s_{3}(u)$ is regular in region 3,
except for the point singularity. The reason for this regularity of
$s_{3}(u)$ is the following: If it was not regular in region 3, we
would have a singularity in the same $u$ on the curve $R=R_{b}$
(Fig. $\ref{fig13}$). This, of course, contradicts our previous
results about the regularity of $s$ in region 2. Therefore the only
point where $s_{3}(u)$ might diverge is the point singularity.
\begin{figure}[htb]
 \vspace{0.5cm}
\epsfxsize=0.6\textwidth
\centerline{\epsffile{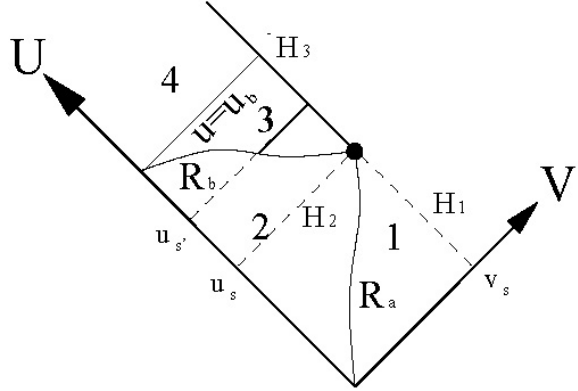}} \vspace*{0.3cm} \caption{\small
\bf \sl A situation where there is a singularity in $s_{3}(u)$ in
$u_{s'} \neq u_{s}$. Then we would have a singularity on $R=R_{a}$
at a regularity point of  $s_{3}$. } \label{fig13}
\end{figure}

b. {}From the inequality for $R_{,u}$ ($\ref{inequa}$), which is
valid in all region 2, we obtain the asymptotic behavior $R_{,u}\sim
e^{k_{2}t}$. Substituting the asymptotic expressions of $R_{,u}$ and
$e^{s_{2}(v)}$ into eq. ($\ref{dvdt}$), we find that $\dot{v} \sim
e^{-k_{1}t}$. Let us denote by $\Delta v$ the distance to the
singularity on the $v$ axes. Then, asymptotically, $\Delta v \sim
e^{-k_{1}t}$ near the singularity ($t \rightarrow \infty$).
Substituting the asymptotic expressions of $R_{,u}$, $e^{s_{2}(v)}$
and $\dot{v}$ into eq. ($\ref{ss3}$) we obtain:
\begin{equation}
e^{s_{4}} \sim e^{(k_{3}-k_{1})t}.
\end{equation}
Substituting $\Delta v$, we obtain:
\begin{equation}
e^{s_{4}}\sim \Delta v^{\frac{k_{3}}{k_{1}}-1} \label{asV}
\end{equation} in
region 4. Equating the expression for $\dot{u}$ in (\ref{general
charac new}) with (\ref{dudt}) and substituting the asymptotic
behavior (\ref{asV})we obtain \be e^{-s}\,R_{,v} \sim \int^{v}\!
\(\Delta
v'\)^{1-\frac{k_{3}}{k_{1}}}\,R_{vv}(0,v')\,dv'-\int^{u}\!h'(R(u',0))\,du'.
\ee Therefore $e^{-s}R_{,v}$ diverges near the inner horizon
$H_{3}$ if $k_{3}>2k_{1}$ and converges if $k_{3}<2k_{1}$.

 c. Recall the evolution equation for $R(u,v)$ in region 4:
\[R_{,uv}=k_{3}e^{s_{4}(v)}.
\]
Integrating this equation with respect to $v$ we obtain:
\begin{equation}
R_{,u}=k_{3}\int^{v}\!e^{s_{4}(v')}dv'+R_{,u}(u,0).
\end{equation}
Substituting the asymptotic expression for $e^{s_{4}}$
($\ref{asV}$) into the expression for $R_{,u}$, we find that at
the inner horizon $R_{,u}$ converges to a constant different than
zero. Therefore the divergences of $e^{-s}R_{,v}R_{,u}$ and
$e^{-s}R_{,v}$ at the inner horizon are the same. $\Box$

\subsection{The No-Hair Phenomenon}

Our semi-linear system demonstrate a no-hair phenomenon.
Asymptotically near the singularity the contribution of the
perturbations disappear. The best way to see it is to look at the
equations and the asymptotic form of their components expressed with
the parameter $t$. This parameter is appropriate for this purpose as
well since $t\rightarrow\infty$ when we expect the perturbations to
vanish. Considering (\ref{the4}) and noting that the equation for
$s$ can be approximated by $\dot{s}\approx k_1+k_2$, we obtain  the
asymptotic forms of $e^{s}$ and $R_{,u}$ in region 2 as $ R_{,u}
\sim e^{k_{2}t}$, $e^{s} \sim e^{(k_{1}+k_{2})t}.$  Substituting
these asymptotic forms in the perturbative term in the first
equation of (\ref{the4}) we obtain
\[
R_{,vv}(0,v)e^{-s}R_{,u} \sim R_{,vv}(0,v)e^{-k_{1}t}.
\]
Thus, when $t\rightarrow\infty$, the contribution of the
perturbation decreases exponentially. Thus the asymptotic form of
the equation for $R_{,v}$ is $\dot{R}_{,v}\sim k_{1}R_{,v}$.
Therefore the asymptotic form of $R_{,v}$ is $R_{,v} \sim
e^{k_{1}t}$ . Substituting the latter in the second equation of
(\ref{the4})  we find that:
\[R_{,uu}(u,0)R_{,v}e^{-s} \sim R_{,uu}(u,0)e^{-k_{2}t}.
\]
Hence, when $t\rightarrow\infty$, the effect of the perturbations
on the equations (and their solution) vanishes. This is the
no-hair phenomenon - the characters of the black hole do not
depend on the explicit form of the perturbations.

\section{Conclusions}

The main goal of this research was to provide evidence for the
validity of the conjecture, which was described in section \ref{the
discussion} (and in~\cite{first}). We actually proved the conjecture
for the particular case of $h(R)$ as a saw-tooth function under few
assumptions. The saw-tooth function is assumed to be generic enough,
so any other function $h(R)$ will yield a similar structure of
singularity for an open neighborhood of the linear case in the space
of initial data.

For $h(R)$ a saw-tooth function our semi-linear system demonstrates
a finite time blowup in region 2 under the following assumptions:
\begin{itemize}
\item $m_{1}<R_{,v}(0,v)<M_{1}$ and $-M_{2}<R_{,u}(u,0)<-m_{2}$
where $m_{1},M_{1},m_{2},M_{2}$ are arbitrary positive constants.
\item $R_{,vv}(0,v)<0$ (the weak energy condition). \item
$R_{a}-\frac{M_{2}R_{,v}(0,0)}{k_{2}}>R_{b}$.
\end{itemize}
In addition our semi-linear system demonstrates a no-hair phenomenon
in region 2. We proved that the singularity in region 2 is actually
a point singularity (which corresponds to future timelike infinity).
The asymptotic expressions near the singular lines in regions 3 and
4 are in full agreement with the conjectured asymptotic behavior
(Table 5.1):
\begin{itemize}
\item $e^{s}\sim (\Delta v)^{\frac{k_{3}}{k_{1}}-1}$. \item
$e^{-s}R_{,v}\sim (\Delta v)^{2-\frac{k_{3}}{k_{1}}}$. \item
$e^{-s}R_{,v}R_{,u} \sim (\Delta v)^{2-\frac{k_{3}}{k_{1}}}$.
\end{itemize}
The ``half gauge invariant scalar''  $e^{-s}R_{,v}$ is related to
the divergent flux at the inner horizon in various models
(see~\cite{first}). The gauge invariant scalar $e^{-s}R_{,v}R_{,u}$
is related to the mass function and the scalar curvature singularity
which is formed when the model contains two intersecting streams of
null fluid.

The curve $R=R_{h}$ (the curve that starts at the point
$u_{h}=k_{2}^{-1}\,R_{,v}(0,0)$ on the $u$ axis) has turned out to
be very important in the definition of region 2 since it is the
first curve in the region which is monotonously decreasing (see Fig.
$\ref{fig11}$). This curve has an important physical meaning for our
model: It is the first characteristic curve ($R=const$) which is
completely inside the trapped region. Namely, $R_{,v}, R_{,u}<0$
along this curve. The apparent horizon is defined as the outer
boundary of the trapped region inside the black hole. In the linear
case (which corresponds to the structure of a static black hole) the
event horizon and the apparent horizon coincide (see Fig.
$\ref{fig4}$). In the non-linear case the apparent horizon is inside
the black hole. In Fig. $\ref{fig11}$ we can see how the apparent
horizon passes if we join the extremal points of the curves behind
the horizon $H_{2}$ by a line - the parts of the curves which are
monotonously decreasing are inside the trapped region.

In conclusion, we found that our semi-linear system with $h(R)$ as a
saw-tooth function demonstrates many features that have physical
meaning  and are analogous to the black-hole inner structure (the
causal structure and location of the singularity, the infinite blue
shift and the inner horizon, mass inflation, the trapped regions
etc.). We even found in the model features of the spacetime outside
the black-hole (e.g. no-hair principle). Further research is needed
to confirm (or contradict) the conjecture under general data. Even
more challenging question is  the validity of the conjecture for a
general $h(R)$.

\bigskip
\noindent{\bf Acknowledgements:} It is a pleasure to thank A. Ori
for the suggestion of the research and for many invaluable
discussions. In addition we thank J. Feinberg for comments on a
preliminary version of the manuscript.

\end{document}